\documentstyle[12pt, psfig]{article}
\def\reference{\parskip 0pt\par\noindent\hangindent 0.5 truecm}

\begin{document}
%
%
\title{Dense Plasma Torus in the GPS Galaxy NGC 1052}
%


\author{Seiji Kameno, $^{1}$ 
 Makoto Inoue, $^{1}$ 
 Kiyoaki Wajima, $^{1}$
 Satoko Sawada-Satoh $^{2}$, \and
 Zhi-Qiang Shen $^{2}$
} 

\date{}
\maketitle

{\center
$^1$ National Astronomical Observatory of Japan, 2-21-1 Osawa, Mitaka Tokyo, Japan, 181-8588\\kameno@hotaka.mtk.nao.ac.jp\\[3mm]
$^2$ The Institute of Space and Astronautical Science, 3-1-1 Yoshinodai, Sagamihara Kanagawa, Japan, 229-8510
}

%
\begin{abstract}
We report results from nearly simultaneous pentachromatic VLBI observations towards a nearby GPS galaxy NGC 1052.
The observations at 1.6 and 4.8 GHz with VSOP, and at 2.3, 8.4, and 15.4 GHz with VLBA, provide linear resolutions of $\sim 0.1$ pc.
Convex spectra of a double-sided jet imply that synchrotron emission is obscured through foreground cold dense plasma, in terms of free--free absorption (FFA).
We found a central condensation of the plasma which covers about 0.1 pc and 1 pc of the approaching and receding jets, respectively.
A simple model with a geometrically thick plasma torus perpendicular to the jets is established to explain the asymmetric distribution of FFA opacities. 

\end{abstract}


\bigskip

%
%

\section{Introduction}
GHz-Peaked Spectrum (GPS) sources show a convex radio spectrum peaked at GHz frequency, as are named (e.g. O'Dea 1998).
The spectral shape indicates that a power-law spectrum of optically thin synchrotron radiation is affected by low-frequency cutoff.
It is a controversial issue what is origin of the low-frequency cutoff; synchrotron self-absorption (SSA) or free--free absorption (FFA).

Tight correlation between the peak frequency and the overall linear size (O'Dea \& Baum 1997), and correlation between the equipartition size and the overall linear size (Snellen et al. 2000) indicates that SSA controls the the peak frequency, at which the equipartition size is derived by Scott \& Readhead (1977).

Bicknell et al. (1997), on the other hand,  pointed out that FFA through ionized gas surrounding radio lobes can also produce the correlation between the peak frequency and the overall size. 
Evidence for FFA was found via multi-frequency VLBI observations in particular GPS sources; 1946+708 (Peck et al. 1999), OQ 208 (Kameno et al. 2000; Xiang et al. 2002), 0108+388 (Marr et al. 2001), and NGC 1052 (Kameno et al. 2001).

We propose a clear method to discriminate the absorption mechanisms.
In the case of SSA, the absorber is the synchrotron emitter itself such as jets or lobes.
Sub-relativistic speeds of the emitter will cause significant Doppler shift of the peak frequency; approaching and receding jets will shows blue- and red-shifted spectra, respectively.
Hence, we expect a higher peak frequency for an approaching jet.
In the case of FFA, ambient cold dense plasma is an absorber whose speed is non-relativistic. 
We expect greater opacity towards receding jets, no matter how the absorber distributes, to produce a higher peak frequency.
This picture was drawn towards some radio galaxies such as 3C 84 (Vermeulen et al. 1994; Walker et al. 1994; Walker et al. 2000), Cen A (Jones et al. 1996; Tingay \& Murphy 2001), NGC 4261 (Jones et al. 2000; Jones et al. 2001), and NGC 6251 (Sudou et al. 2000).
The opposite spectral behavior between SSA and FFA allows us to discriminate the absorption mechanisms by measurements of the peak frequencies towards approaching and receding jets.
Multi-frequency observations towards GPS sources with double-sided jets are relevant to this issue.

We chose the nearest GPS source, NGC 1052, for this study.
The redshift of this galaxy, $z=0.0049$ (Knapp et al. 1978),  corresponds to the distance of 20 Mpc, if we assume $H_0 = 75$ km s$^{-1}$ and $q_0 = 0.5$, and 1 milliarcsec (mas) corresponds to 0.1 pc.
This object has parsec-scale twin jets in P.A. $\sim 65^{\circ}$ (Claussen et al. 1998), so that it is suitable for the discrimination of the absorbing mechanism.
Their observations also revealed that H$_2$O maser spots distribute along the western jet with a velocity gradient.

Our previous trichromatic VLBA observations for the first time clarified the jet geometry and presence of a dense plasma torus that is responsible for FFA (Kameno et al. 2001).
The eastern and western sides of the jet are approaching and receding with the velocity of $\beta = 0.25$ at the viewing angle of $50^{\circ}$.  
The western receding jet is obscured at lower frequencies, suggesting that a dense plasma torus covers the receding jet to produce FFA.
Although the results were successful, three frequencies were not sufficient to obtain confident fit for the spectrum.
 
In this paper, we report new pentachromatic VLBI observations of NGC 1052, using the VSOP (Hirabayashi et al. 1998) and the VLBA.

\begin{figure}[htb]
\centerline{
\psfig{file=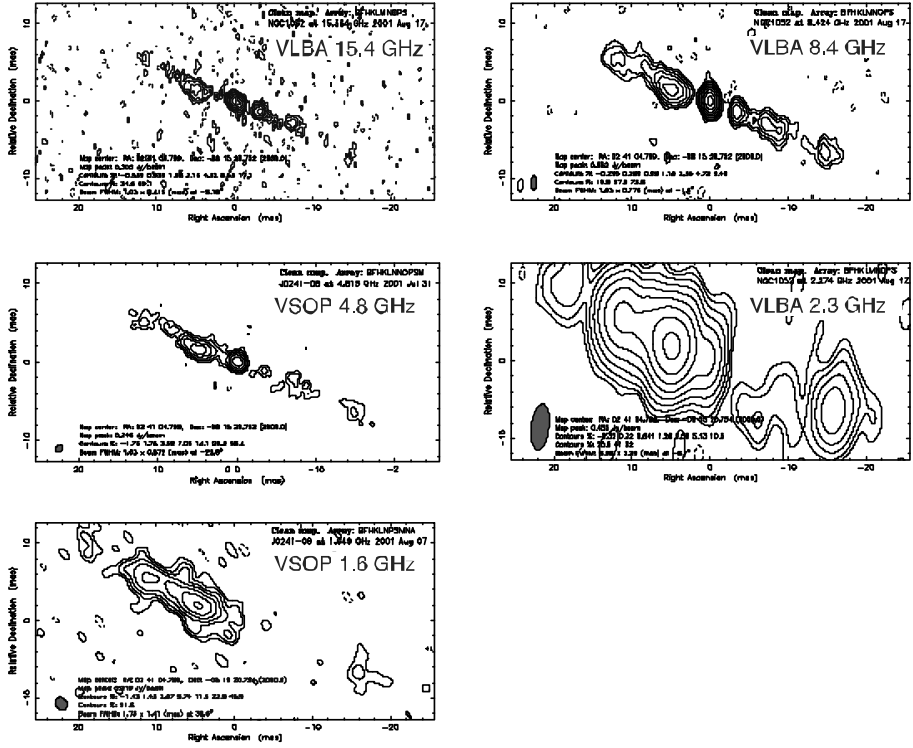,width=120mm}
}
\caption{Pentachromatic images of NGC 1052. Image performance is listed in table \ref{tab:imagespec}. Contours start at $\pm 3$ times the r.m.s. level and increase by factors of 2. Synthesized beams are shown at left bottom corner in each image.}
\label{fig:pentachromat}            
\end{figure}

\begin{figure}[htb]
\centerline{
\psfig{file=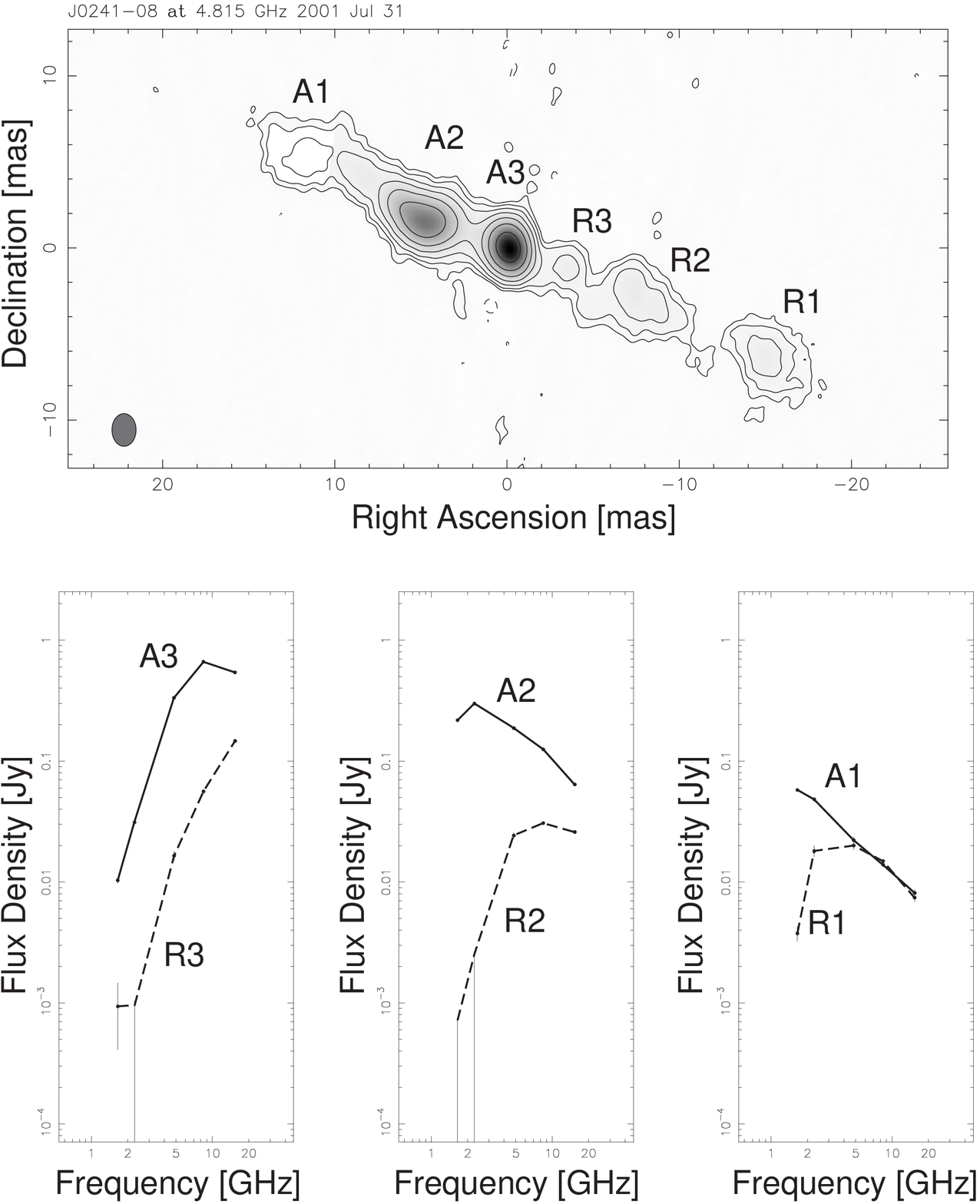,width=90mm}
}
\caption{Top: Total intensity map at 4.8 GHz convolved with an elliptical Gaussian whose FWHM is $1.9 \times 1.4$ mas in p.a.$=0^{\circ}$.  Contours start at $\pm 3\sigma$ level and increase by factors of 2, where $\sigma = 1.13$ mJy beam$^{-1}$. Approaching and receding components are marked as A$_k$ and R$_k$, where $k=1$, 2, and 3. Bottom: Spectra of components.}
\label{fig:contspec}            
\end{figure}

\section{Observations and Results}
All pentachromatic observations have been carried out within 18 days during July 31 through August 31, 2001.
We used the space-VLBI VSOP array at 1.6 and 4.8 GHz, and VLBA at 2.3, 8.4, and 15.4 GHz.
This combination provides comparable resolutions of $\sim 1$ mas through an order of frequency magnitude except at 2.3 GHz.

Observations at 2.3, 8.4, and 15.4 GHz using VLBA, coded BK084, were done on August 7, 2001.
We used the dual-frequency 2.3/8.4 and the 15.4 GHz receiving systems, switching by 22-min scan cycles.
Total integration time consists of 10 and 11 scans for 2.3/8.4 and 15.4 GHz, respectively.
Four 8-MHz baseband converters (BBCs) are allocated for 15.4 GHz, while 2 BBCs are used for 2.3 and 8.4 GHz, respectively.
Fringes were detected with sufficient signal-to-noise ratios (SNRs) for all valid baselines and scans. 

Space-VLBI VSOP observations at 1.6 GHz (W513A) and 4.8 GHz (W513B) were executed on August 7 and July 31, 2001, respectively.
The spacecraft HALCA was linked to two tracking stations during each observation, Green Bank and Tidbinbilla for W513A, and Green Bank and Usuda for W513B.
Though fringes within ground array were strong enough, those in space baselines were marginal at 4.8 GHz.
While visibilities with the tracking pass using Green Bank were available, we missed them in the Usuda tracking pass.
Fringes at 1.6 GHz were entirely detected within ground and space baselines.

Fringe fitting, passband correction, and {\it a priori} amplitude calibrations were taken using the NRAO AIPS.
Image synthesis and self calibration were processed with Difmap. 
Image performance is listed in table \ref{tab:imagespec}.

\footnotesize
\begin{table}[ht]
\caption{Image performance}
\begin{tabular}{rrrrrl} \hline \hline
\multicolumn{1}{c}{Frequency} &
\multicolumn{3}{c}{Beam} &
\multicolumn{1}{c}{Image r.m.s.} & 
\multicolumn{1}{c}{Telescopes}  \\ \cline{2-4}
\multicolumn{1}{c}{(GHz)} &
\multicolumn{1}{c}{Major (mas)} &
\multicolumn{1}{c}{Minor (mas)} &
\multicolumn{1}{c}{P.A. (deg.)} &
\multicolumn{1}{c}{(mJy beam$^{-1}$)} \\ \hline
1.6   & 1.75 & 1.39  & $31.3$  & 1.183 & HALCA, ATCA, and VLBA \\
2.3   & 6.34 & 2.58  & $-5.8$  & 0.511 & VLBA \\
4.8   & 0.78 & 0.67  & $-78.7$ & 1.473 & HALCA and VLBA \\
8.4   & 1.86 & 0.77  & $-2.2$  & 0.444 & VLBA \\
15.4  & 1.09 & 0.42  & $-7.8$  & 0.432 & VLBA \\ \hline
\end{tabular}
\label{tab:imagespec}
\end{table}
\normalsize

Figure \ref{fig:pentachromat} shows images at five frequencies.
Each image includes uncertainty in absolute position through self-calibration process.
To register the images at different frequencies, we picked up 6 distinct components; A1, A2, and A3 in the approaching jet, and R1, R2, and R3 in the receding jet.
We then measured positions $(\xi_{\rm k}^{\nu}, \eta_{\rm k}^{\nu})$ of them with respect to a tentative origin of each image.
Here k and $\nu$ stand for indices of components and frequency, respectively.
We then derived relative offsets $(\delta \xi^{\nu}, \delta \eta^{\nu})$ to minimize the positional residuals $\chi_{\nu}^2$ defined as
\begin{eqnarray}
\chi_{\nu}^2 = \sum_{\rm k} \left[ \frac{(\xi_{\rm k}^{\nu} - \delta \xi^{\nu} - \xi_{\rm k}^{\nu_0})^2}{\sigma_{\xi_{\rm k}^{\nu}}^2 + \sigma_{\xi_{\rm k}^{\nu_0}}^2} + \frac{(\eta_{\rm k}^{\nu} - \delta \eta^{\nu} - \eta_{\rm k}^{\nu_0})^2}{\sigma_{\eta_{\rm k}^{\nu}}^2 + \sigma_{\eta_{\rm k}^{\nu_0}}^2} \right], 
\end{eqnarray}
where $\sigma_{\xi_{\rm k}^{\nu}}$ and $\sigma_{\eta_{\rm k}^{\nu}}$ are standard positional errors of component k at frequency $\nu$, and $\nu_0$ is the frequency of the reference image. 
We take $\nu_0 = 4.8$ GHz, where the frequency is middle of our observing range, and where the synthesized beam is the narrowest.
Standard errors of the registration are listed in table \ref{tab:registerror}.

\footnotesize
\begin{table}[ht]
\caption{Accuracy of image registration}
\begin{center}
\begin{tabular}{rrr} \hline \hline
\multicolumn{1}{c}{Frequency} &
\multicolumn{1}{c}{R. A.} &
\multicolumn{1}{c}{Dec.} \\
\multicolumn{1}{c}{(GHz)} &
\multicolumn{1}{c}{(mas)} &
\multicolumn{1}{c}{(mas)} \\ \hline
15.4  & 0.30  & 0.29 \\
8.4   & 0.21  & 0.31 \\
4.8   & --    & --  \\
2.3   & 2.80  & 0.75 \\
1.6   & 0.94  & 0.43 \\ \hline
\end{tabular}
\end{center}
\label{tab:registerror}
\end{table}
\normalsize

\begin{figure}[htb]
\centerline{
\psfig{file=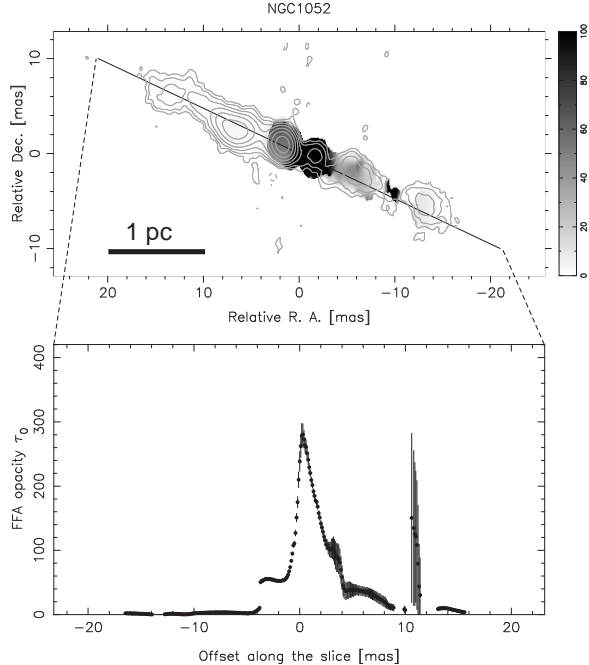,width=80mm}
}
\caption{Top: Spatial distribution of FFA opacity $\tau_0$ in grey scale overlaid to the total intensity map at 4.8 GHz in contour. Bottom: Profile of the FFA opacity along the jet. Errors are derived from the nonlinear spectral fit using equation \ref{eqn:ffamodelfit}.}
\label{fig:grey+slice}            
\end{figure}

\section{Discussion}
First of all, we aim at clarifying the absorbing mechanism.
Figure \ref{fig:contspec} shows continuum spectra of 6 components.
Receding components R1, R2, and R3 show sharp cutoff at low-frequency ends.
The spectral index of R2 between 1.6 and 2.3 GHz is $\alpha_{1.6}^{2.3} = 4.76 \pm 0.48$.
Here we define the spectral index $\alpha$ as $S_{\nu} \propto \nu^{+\alpha}$, where $S_{\nu}$ is the flux density and $\nu$ is the frequency.
The spectral index exceeds maximum attainable spectral index of 2.5 by SSA.
For components R2 and R3 $\alpha_{1.6}^{4.8} > 2.56$ and $\alpha_{1.6}^{4.8} = 2.43 \pm 0.38$, respectively, are close to the SSA limit.
We evaluated the spectral indices between 1.6 and 4.8 GHz for these components because flux densities at 2.3 GHz is too weak to detect.
The brightest approaching component, A3, shows $\alpha_{1.6}^{4.8} = 3.28 \pm 0.05$.
These steeply rising spectra indicate significant presence of FFA, which can produce $\alpha > 2.5$.

Component A1 shows a single power-law spectrum with $\alpha = -1.1$, whose spectral peak should be at $< 1.6$ GHz, while the spectral peak of its counter component R1 is between 2.3 and 4.8 GHz. 
Components A2 and A3 mark the spectral peaks at $\sim 2.3$ and $\sim 8.4$ GHz, respectively, whose counter components R2 and R3 show the spectral peaks at $\sim 8.4$ and $> 15.4$ GHz.
Every approaching component shows a lower peak frequency than each counterpart does.
As is mentioned in the introduction, this manner clearly indicates that the receding jet is obscured by external absorber, say, FFA.

\begin{figure}[htb]
\centerline{
\psfig{file=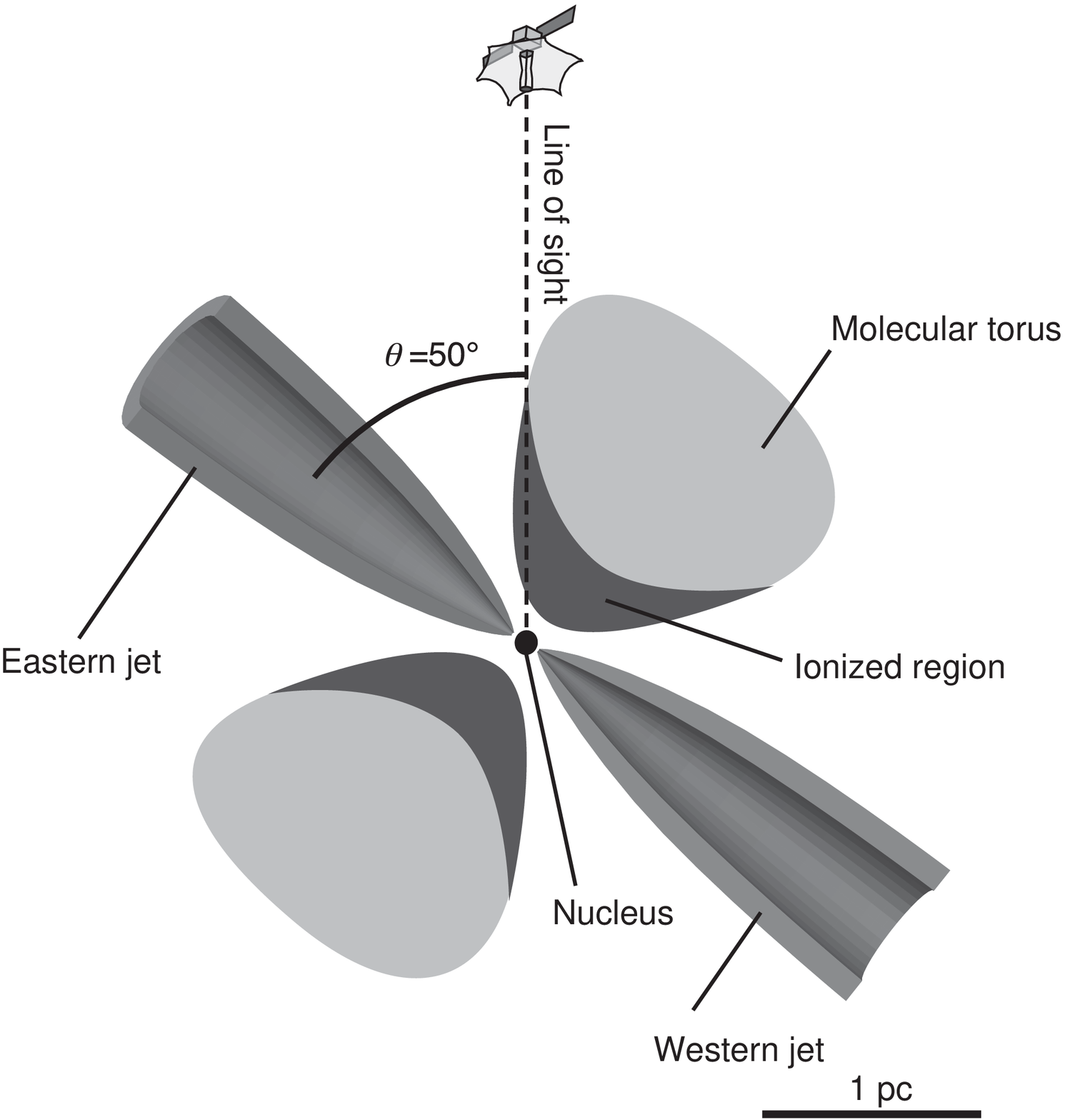,width=80mm}
}
\caption{Schematic diagram of NGC 1052. Double-sided jet is inclined by $50^{\circ}$ with respect to the line of sight. A geometrically thick molecular torus is perpendicular to the jet, whose inner surface is illuminated by ionizing photons from the nucleus. The ionized region obscures radio emission from the jet, in terms of free--free absorption. Long path length through the ionized region towards the nucleus results in a large opacity, as is shown in figure \ref{fig:grey+slice}}
\label{fig:torusmodel}            
\end{figure}

To illustrate spatial distribution of opacity, we applied spectral fit for the FFA model
\begin{eqnarray}
S_{\nu} = S_0 \nu^{\alpha_0} \exp( -\tau_0 \nu^{-2.1} ), \label{eqn:ffamodelfit}
\end{eqnarray}
pixel by pixel.
Here, $S_{\nu}$ is the observed flux density at the frequency $\nu$ in GHz, $\alpha_0$ is the intrinsic spectral index, $S_0$ is the absorption-corrected flux density at 1 GHz, and $\tau_0$ is the FFA coefficient. 
We fixed $\alpha_0$ to $-1.1$, which is obtained from the spectral index of component A1.
Figure \ref{fig:grey+slice} shows spatial distribution of $\tau_0$ and its profile along the jet.
The maximum value of $\tau_0 \sim 300$ is consistent with Kameno et al. (2001).
The profile tails over $\sim 1$ pc towards the western receding jet, while it rapidly decreases towards the eastern approaching jet.
This asymmetric profile is again consistent with Kameno et al. (2001), with more detailed structure at the hands of high resolution by VSOP.  

Since the opacity coefficient, $\tau_0$, is related to the electron density, $n_{\rm e}$, and temperature, $T_{\rm e}$, along line of sight as
\begin{eqnarray} 
\tau_0 = 0.46 \int_{\rm LOS} n_{\rm e}^2 T_{\rm e}^{-\frac{3}{2}} dL,
\end{eqnarray} 
it will be an indicator of cold dense plasma.
Here, the integration stands for path length along the line of sight in unit of pc.
When the aborber is homogeneous, $\tau_0$ is proportional to the path length through the ionized region.
A simple model to explain the opacity profile is shown in figure \ref{fig:torusmodel} with a geometrically thick torus perpendicular to the jets.
The inner surface of the torus is photo-ionized by illumination of ionizing photons from the nucleus, while outer region remains in neutral gas or dusts.
The path length through the ionized region can produce the asymmetric profile of the opacity, when we look from the viewing angle of $50^{\circ}$. 

We can estimate physical condition of the absorber, if we assume the torus model.
The maximum value of $\tau_0 \sim 300$ towards nucleus gives 
\begin{eqnarray} 
n_{\rm e}^2 T_{\rm e}^{-\frac{3}{2}} = 650,
\end{eqnarray} 
when the path length is the same the width of the absorbed area along the receding jet, say, $L \sim 1$ pc.
The ionizing condition of $T_{\rm e} \ge 10^4$ K gives the lower limit of $n_{\rm e} \ge 2.5 \times 10^4$ cm$^{-3}$.
The upper limit of $T_{\rm e}$ can be given by absence of significant free--free emission below the brightness temperature of $T_{\rm b} \leq 4.9 \times 10^6$ K at 15.4 GHz or $T_{\rm b} \leq 5.4 \times 10^6$ K at 8.4 GHz.
Since the torus is optically thick, in terms of bremsstrahlung, we have $T_{\rm e} \sim T_{\rm b} \leq 4.9 \times 10^6$ K and then $n_{\rm e} \leq 2.6 \times 10^6$ cm$^{-3}$.  
This evaluation indicates that the FFA absorber consists of cold ($10^4 \leq T_{\rm e} \leq 4.9 \times 10^6$ K) and dense ($2.5 \times 10^4 \leq n_{\rm e} \leq 2.6 \times 10^6$ cm $^{-3}$) plasma.
The average value of electron density derives the electron column density across the plasma torus ($L \sim 1$ pc),
\begin{eqnarray} 
n_{\rm e} L = 0.8 \times 10^{23} \left( \frac{T_{\rm e}}{10^4 \: {\rm K}} \right)^{\frac{3}{4}} \: \mbox{cm}^{-2}.
\end{eqnarray} 
This electron column density is comparable to the atomic column density of $\sim 10^{23}$ cm$^{-2}$ measured by ROSAT and ASCA X-ray observations (Guainazzi \& Antonelli 1999).

We mention that the location of water maser spots (Claussen et al. 1998) coincides with where the FFA opacity is large. 
The torus model explains results of the water masers; the H$_2$O molecules survive inside the torus, shielded against exposure by the nucleus.
The seed photon from the receding jet are amplified by excited molecules in the torus.
Amplification does not work towards the approaching jet, because the torus is not in front of the jet, the seed photon supplier.
Absence of masers towards the eastern jet can be understood by this model.
A detailed monitoring program would verify this model by detection of orbital motion of maser spots within the torus.


%
%




\section*{Acknowledgments}

The authors gratefully acknowledge the VSOP Project, which is led by the Japanese Institute of Space and Astronautical Science in cooperation with many organizations and radio telescopes around the world.
We the VLBA operated by the National Radio Astronomy Observatory, which is a facility of the National Science Foundation (NSF).
This work was supported by Grant in Aid (c) 13640248 from Japan Society for the Promotion of Science.

\section*{References}






\reference Bicknell, G. V., Dopita, M. A., \& O'Dea, C. P. \  1997, ApJ, 485, 112

\reference Claussen, M. J., Diamond, P. J., Braatz, J. A., Wilson, A. S., \& Henkel, C. \  1998, ApJ, 500, L129

\reference Guainazzi, M., \& Antonelli, L. A. \ 1999, MNRAS, 304, L15

\reference Hirabayashi, H., et al. \ 1998, Science, 281, 1825

\reference Jones, D. L., et al. \ 1996, ApJ, 466, 63

\reference Jones, D. L., Wehrle, A. W., Meier, D. L., \& Piner, B. G. \  2000, ApJ, 534, 165

\reference Jones, D. L., Wehrle, A. W., Piner, B. G. \& Meier, D. L. \  2001, ApJ, 553, 968

\reference Kameno, S, Horiuchi, S., Shen, Z.-Q., Inoue, M., Kobayashi, H., Hirabayashi, H., \& Murata, Y. \  2000, PASJ, 52, 209

\reference Kameno, S, Sawada-Satoh, S., Inoue, M., Shen, Z.-Q., \& Wajima, K. \  2001, PASJ, 53, 169

\reference Marr, J. M., Taylor, G. B., \& Crawford, F. III \  2001, ApJ, 550, 160

\reference O'Dea C. P. \  1998, PASP, 110, 493

\reference O'Dea C. P., \& Baum, S. A. \  1997, AJ, 113, 148

\reference Peck, A. B., Taylor, G. B., \& Conway, J. E. \  1999, ApJ, 521, 103

\reference Scott, M. A., \& Readhead, A. C. S. \  1977, MNRAS, 180, 539

\reference Snellen, I. A. G., Schillizi, R. T., Miley, G. K., de Bruyn, A. G., Bremer, M. N., \& R\"ottgering, H. J. A. \  2000, MNRAS, 319, 445

\reference Sudou, H., Taniguchi, Y., Ohyama, Y., Kameno, S., Sawada-Satoh, S., Inoue, M., Kaburaki, O., \& Sasao, T. \  2000, PASJ, 52, 989

\reference Tingay, S. J., \& Murphy, D. W. \  2001, ApJ, 546, 210

\reference Vermeulen, R. C., Readhead, A. C. C., \& Backer, D. C. \  1994, ApJ, 430, L41

\reference Walker, R. C., Dhawan, V., Romney, J. D., Kellermann, K. I. \& Vermeulen, R. C. \  2000, ApJ, 530, 233

\reference Walker, R. C., Romney, J. D., \& Benson, J. M. \  1994, ApJ, 430, L45

\reference Xiang, L., Stanghellini, C., Dallacassa, D., \& Haiyan, Z. \  2002, A\&A, 385, 768


\end{document}